\documentclass[12pt]{iopart}
\usepackage{iopams} 
\input epsf
\expandafter\let\csname equation*\endcsname\relax
 \expandafter\let\csname endequation*\endcsname\relax
 \usepackage[fleqn]{amsmath}
\usepackage{amsopn}
\usepackage{epsfig}
\usepackage{subfig}
\usepackage{graphics,psfrag,rotating}
\usepackage{graphicx}
\usepackage{dcolumn}
\usepackage{bm}
\usepackage{epstopdf}
\usepackage{color}
\usepackage[usenames,dvipsnames,svgnames]{xcolor}
\usepackage[colorlinks=true,
      linkcolor=red,
      urlcolor=gray,
      citecolor=blue]{hyperref}
 \usepackage[]{cite}

\def\3nab{\tilde{\nabla}}

\def\hsp5{\hspace{5mm}}

\def\case#1/#2{\textstyle\frac{#1}{#2}}

\def\ber {\begin{eqnarray}}
\def\eer {\end{eqnarray}}
\def\bea {\begin{eqnarray}}
\def\eea {\end{eqnarray}}

\def\bc {\begin{center}}
\def\ec {\end{center}}
\def\case#1/#2{\frac{#1}{#2}}

\newcommand{\be}{\begin{equation}}
\newcommand{\bse}{\begin{subequation}}
\newcommand{\ese}{\end{subequation}}
\newcommand{\ee}{\end{equation}}
\newcommand{\eei}{\end{eqnarray}\indent\indent}
\newcommand{\ba}{\begin{array}}
\newcommand{\ea}{\end{array}}
\newcommand{\bal}{\begin{eqnarray}}
\newcommand{\eal}{\end{eqnarray}}

\def\case#1/#2{\textstyle\frac{#1}{#2} }

\newcommand{\bver}{\begin{verbatim}}
\newcommand{\ever}{\end{verbatim}}

\begin{document}
\title{Linear Cosmological perturbations in $f(Q)$ Gravity}
\author{ Shambel Sahlu$^{1,2}$, Endalkachew Tsegaye$^2$}
\address{$^{1}$ Astronomy and Astrophysics Research and Development Department,\\ Entoto Observatory and Research Centre.}
\address{$^{2}$ Department of Physics, College of Natural and Computational Science, Wolkite University, Wolkite, Ethiopia}
\date{\today}

\begin{abstract}
In this manuscript, we studied the accelerated expansion history of the universe and the formations of large-scale structures using $f(Q)$ gravity model. The expansion rate of the universe within distance modulus and redshift has been performed in $f(Q)$ gravity. This work is also devoted to investigating the linear cosmological perturbations in the $f(Q)$ gravity model using the $1 + 3$ covariant formalism. The scalar and harmonic decomposition techniques are applied to find the evolution equation. The growth of matter density contrasts is analyzed with redshift $z$. Using the nonmetricity modified gravity model which is $f (Q) = Q+\alpha Q^n$, the accelerating expanding universe and the formations of the large-scale structures have been explained. The quasi-static approximation technique is applied to analyze the growth of matter contrasts. The growth of density contrast is studied within the dust and radiation-dominated universe in the presence of nonmetricity gravity.
\begin{description}
\item[PACS numbers]{04.50.Kd, 98.80.Jk, 98.80.-k, 95.36.+x, 98.80.Cq}\\
\end{description}
\end{abstract}

\section{Introduction}
The modification of General Relativity (GR) \cite{elmardi2016chaplygin,sahlu2019chaplygin,benaoum2012modified,sahlu2019accelerating} has been gained much attention to explain the late-time cosmic accelerated expansion history of the universe\cite{linder2003exploring, harko2018coupling,haridasu2018improved} without the need for the dark-energy scenario.The symmetric teleparallel gravity so called  $f(Q)$ gravity theory is  one of a modified theory gravity which has vanishing curvature  and 	torsion connection. $f(Q)$  gravity is described by a nonmetricity scalar $Q$ and it represents one of the geometrical equivalent formulations of GR\cite{jimenez2020cosmology, khyllep2021cosmological,solanki2021cosmic}.
Some bservational constraints  has been investigated in $f(Q)$ model  using  SNIa data.  And the $f(Q)$ model is a good candidate as $\Lambda$CDM to describes the accelerated expansion of universe.
\\
\\
The cosmological solutions and growth index of matter perturbations in $f(Q)$ gravity has been studied in \cite{khyllep2021cosmological} the metric formalisim approach and here in this  manuscript, we study the linear cosmologica perturbations using 1+3 covariant formalisim as presented in \cite{bruni1992cosmological, 38, ntahompagaze2018study,sami2021covariant} to investigate the evolution of structure and to compute how the gravitational forces causing small perturbations to grow and eventually seed the formation of stars, quasars, galaxies and clusters\cite{fry1984galaxy}. The theory of cosmological perturbations has become a cornerstone of modern  cosmology since it is the framework which provides the link 	between the models of  early Universe such as the inflationary universe scenario which yield causal mechanisms for the generation of density fluctuations and allows us to connect.
\\
\\
This paper mainly focus on to investigate the linear cosmological perturbations in $f(Q)$ gravity model using the 1+3 covariant formalism.  The power law model $f(Q)=Q+\alpha Q^n$ in\cite{khyllep2021cosmological,ayuso2021observational}  is considered  to find the density contrast through cosmic time. For the case of $f(Q)=Q$,  the symmetric teleparallel gravity equivalent to GR and the field equation also reduced to GR.  The growth of density contrast is investigated and clearly observe the contribution of nonmetricity scalar to study the formation of large-scale  as presented in \cite{khyllep2021cosmological,ayuso2021observational, sahlu2020scalar}.
\section{$f(Q)$ Gravity Cosmology}
	Curvetureless and torsion-free connection describes the symmetric teleparallel equivalen of general relativity (STEGR) we call it concident general relativity.
	The covariant form of the field equation of $f(Q)$ gravity generates based on the fundamental object is the nonmetricity
	tensor given by
	\begin{equation}
		Q_{cab}=\triangledown_c g_{ab},
	\end{equation} 
	where $g_{ab}$ is the metric. The field equation for $f(Q)$ gravity obtained from the action presented in\cite{khyllep2021cosmological}.
	We consider a modified gravity
	theory in which the fundamental object is the nonmetricity scalar   given by 
	\begin{equation}
		S=\int\frac{1}{2}f(Q)\sqrt{-g} d^4x+\int L_m \sqrt{-g}d^4x, 
	\end{equation}
	where $f(Q)$ is an arbitrary function of the nonmetricity $Q$, $\sqrt{-g}$ is the determinant of the metric $g_{ab}$ and $L_m$ is the matter lagrangian density. It is also useful to introduce the superpotential
	\begin{equation}
		4P^c_{ab}=-Q^c_{ab}+2Q^c_{(ab)}-Q^cg_{ab}-\tilde{Q}^c g_{ab}-\delta^c_{(a}{(Q_b)}.
	\end{equation}
	Varying the action with respect to the metric,
	one obtains the gravitational field equation given by
	\begin{equation}
		\frac{2}{\sqrt{-g}}\triangledown_c(\sqrt{-g}f'P^c\hspace{0.01cm}_{ab})+\frac{1}{2}g_{ab}f+f_Q(P_{abd}Q_b\hspace{0.01cm}^{cd}-2Q_{cd a}P^{c\beta}\hspace{0.01cm}_b)=-T_{ab}. \label{4}
	\end{equation}
	The field equation in \eqref{4} we can also rewrite as
	\begin{equation}
		f'G_{ab}+\frac{1}{2}g_{ab}(f'Q-f)+2f''\nabla_cQ^c_{ab}=T_{ab},	\label{5}
	\end{equation}
	where $g_{ab}$ is metric, $f''=\frac{d^2f}{dQ^2}$, $f'=\frac{df}{dQ}$ and $T_{ab}=-\frac{2}{\sqrt{-g}}\frac{\delta S_m}{\delta g_{ab}}$ is the matter field of EMT. The energy momentum tensor of a perfect fluid is defind as
	\begin{equation}
		T_{ab}=(\rho+p)u_au_b+pg_{ab},
	\end{equation}
  where $u_a$ is the four-velocity satisfying the normalization
	condition $u_au_b= -1$, $\rho$ and p are the energy density and
	pressure of a perfect fluid respectively.
	Under the homogeneous and isotropic universe described by the
	Friedmann-Lemaitre-Robertson-Walker metric
	\begin{equation}
		ds^2=-dt^2+a^2(t)g_{ab}dx^adx^b,
	\end{equation}
	where a(t) is the scale factor of the universe. \\\\
	From the above modified field equation \eqref{5}, the modified Friedmann equations is presented as\cite{khyllep2021cosmological} for the case of $f(Q)=Q+F(Q)$, and its
	corresponding  thermodynamic quantities  yield as
	Where $\rho$ and p are the density and pressure respectively given by
	\begin{eqnarray}
		&& \rho_{eff} = \rho_m +\rho_Q,\\&&
		p_{eff} = p_m +p_Q.
	\end{eqnarray}
	And the corresponding nonmetricity energy density and pressure become  
	\begin{eqnarray}
		&&\rho_Q=\frac{F}{2}-QF',\label{55}\\&&
		p_Q = -\rho_Q + \frac{\theta \dot{Q}}{3Q}(2QF''+F'). \label{555}
	\end{eqnarray}
	and the modified Friedmann equations given as
	\begin{eqnarray}
		&&H^2=\frac{1}{3}(\rho_m+\frac{F}{2}-QF')\label{33ab},\\&&
		2\dot{H}+3H^2=-\frac{\rho_{eff}}{3}-p_{eff}\label{333}.
	\end{eqnarray}
	From equations  \eqref{55}, \eqref{555}, \eqref{33ab} and \eqref{333} we can write the energy density and pressure of the universe respectively as follow:
	\begin{eqnarray}
		\frac{\dot{H}}{H^2}= -\dfrac{3}{2\mathcal{X}}+\frac{1}{2}\left( -\frac{\Omega_m}{\mathcal{X}}+\frac{\Re}{3\mathcal{X}} \right) \label{19},
	\end{eqnarray}
	where $\mathcal{X} = (1+2QF''+F'),\qquad \Omega_m=\frac{\rho}{3H^2}, \qquad \Re=\frac{F-2QF'}{6H^2}$.
		In this manuscript, we we consider the power-law $F(Q) = \alpha Q ^n$  gravity model, and $\alpha$ becomes
	\begin{equation}
		\alpha = \frac{(1-\Omega_m)6H_0^2}{Q^n(1-n)}.
	\end{equation}
	The normalize Hubble parameter $h(z)$ in the presence of redshift the power-law $f(Q)$ model expressed as
	\begin{eqnarray}
		h(z)=\sqrt{(1+z)^3\Omega_m+\frac{(1-\Omega_m)(1-2n)}{1-n}} \label{19}.
	\end{eqnarray}
	For the case of  $n=0$ equation \eqref{19} reduced to the $\Lambda CDM$ model  as presented \cite{moresco20166}
	\begin{equation}
		h(z)= \sqrt{(1+z)^3\Omega_m+\Omega_\Lambda},\quad \label{17}
	\end{equation}
	where $h=\frac{H}{H_0}$ and $\frac{a_0}{a}=1+z$, $a_0$ is the scale factor at the present time which we can normalize to 1. The best-known way to trace the evolution of the universe observationally is to look into the redshift - luminosity distance relation.  
	Now we can calculate the distance modulus $\mu$ for \eqref{19} is given by
	\begin{equation}
		\mu=m-M=25-5\times \log_{10}\left\lbrace3000\bar{h}^{-1}(1+z)\int_{0}^{z}\frac{dz'}{h(z')} \right\rbrace, \label{24}
	\end{equation}
	where apparent magnitude m and absolute magnitude M are logarithmic measure of flux and luminosity respectively as well as the line-of-sight comoving distance takes a value, $f(z)=3000\bar{h}^{-1}km/sMPc$ and the Hubble uncertainty parameter is given interms of Hubble parameter by $\bar{h}=\frac{H(z)}{100km/sMPc}$.
		\begin{figure}[h!]
		\centering
		\includegraphics[width=12cm, height=8cm]{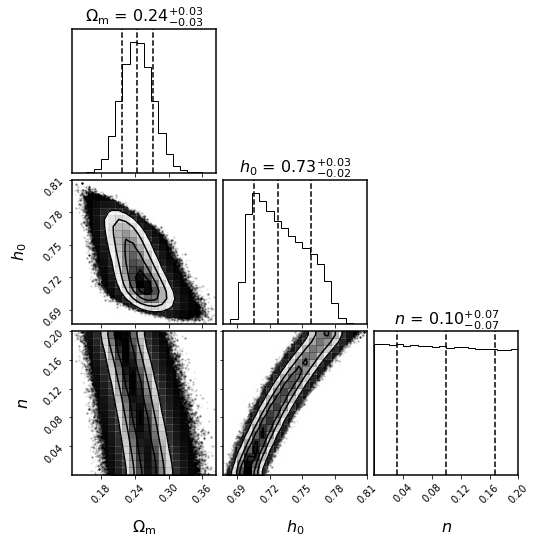}
		\caption{Some cosmological constraints in  $f(Q)$ gravity model using SNIa data. We used 100 random walkers and 50000 iterations.}
		\label{fig:f_Q}
	\end{figure}
Here we us MCMC simulation  fits at the average values of free constrained parameters we had the matter density, $\Omega_m$ and for the Hubble uncertainty parameter, $\bar{h}$ and the free parameter $n$ as presented in Fig. \ref{fig:f_Q}. From the plots we observe that $f(Q)$ gravity to explore the late-history of the universe as the well known model $\Lambda$CDM.
\section{Linear Cosmological Perturbations in $f(Q)$ Gravity}
The Raychaudhuri equation of the fluid for GR limits as presented in \cite{38, ntahompagaze2018study,sami2021covariant,sahlu2020scalar}
	\begin{equation}
		\dot{\theta}=-\frac{1}{3}\theta^2-\frac{1}{2}(\rho+3p)+\nabla^a\dot{u}_a. \label{32}
	\end{equation}
	For the case of $f(Q)$ gravity model Eq. \eqref{32} yields as
	\begin{equation}
		\dot{\theta}=-\frac{1}{3}\theta^2-\frac{1}{2}(1+3w)\rho_m-\frac{1}{2}\left( -2\rho_Q+\frac{\theta\dot{Q}}{Q}(2QF''+F')\right) +\nabla^a\dot{u}_a.
	\end{equation}
	This is a key equation to find the evolution equations. 
	At the cosmological background level we work on a homogeneous and isotropic expanding (FLRW) univerese based on spatial gradients of gauge-invariant variables such as $D^m_a$ and $Z_a$ represents energy density and volume expansion of the fluid respectively are the basic tools to extract the evolution equation for matter density fluctuations shows as 
	\begin{eqnarray}
		&&D^m_a=\frac{a}{\rho_m}\tilde{\nabla}_a\rho_m,\label{35}\\&&
		Z_a=a\tilde{\nabla}_a\theta.\label{36}
	\end{eqnarray}
	Based on the nonmetricity fluid for $f(Q)$ gravity by introducing new terms $\mathcal{W}_a$ and $\mathcal{L}_a$ resulting from spatial gradients of gauge-invariant quantities to characterize the fluctuations in the nonmetricity density and momentum respectively defined as
	\begin{eqnarray}
		&&\mathcal{W}_a=a\tilde{\nabla}_aQ,\label{ggg}\\&&
		\mathcal{L}_a=a\tilde{\nabla}_a\dot{Q}.\label{oooo}
	\end{eqnarray}
	The conservation equations for non interacting fluid are given by
	\begin{eqnarray}
		&&\dot{\rho}_m=-\theta(\rho_m+p_m)+(\rho_m+p_m)\tilde{\nabla}^a\Psi_a\label{w}\\&&
		(\rho_m+p_m)\dot{u}_a+)\tilde{\nabla}_ap_m+\dot{\Psi}_a-(3c^2_s-1)\frac{\theta}{3}\Psi_a+\Pi_a=0  \label{k},	
	\end{eqnarray}
	where 
	\begin{equation}
		\Psi_a=\frac{q_a}{(\rho_m+p_m)},\hspace{1cm}\Pi_a=\frac{\tilde{\nabla}^b\pi_{ab}}{(\rho_m+p_m)}.
	\end{equation}
	And also $\Psi_a$ and $\Pi_a$ is vanish for perfect fluid, in our case we consider the perfect fluid. 
	From equations \eqref{35} and \eqref{36} we can get the first order matter fluid perturbation evolution equations takes a form
	\begin{eqnarray}
		&& \dot{D}^m_a=-(1+w)Z_a+w\theta D^m_a \;, \\
		&&\dot{Z}_a=-\left[ \frac{2\theta}{3}+\frac{\dot{Q}F'}{2Q}-\dot{Q}F''\right]Z_a+\left[ \frac{w\theta^2}{3(1+w)}+\frac{w(1+3w)}{2(1+w)}\rho_m-\frac{w F}{2(1+w)},\nonumber\right.\\&&\left.-\frac{QF'w}{1+w}+\frac{\theta\dot{Q}F''w}{1+w}+\frac{\theta\dot{Q}F'w}{2Q(1+w)}-\frac{1}{2}(1+3w)\rho_m-\tilde{\nabla}^2 \frac{w}{1+w}\right] D^m_a+\left[ \frac{1}{2}F',  \nonumber\right.\\&&\left. -QF''-F'+\theta\dot{Q}F'''-\frac{\theta\dot{Q}F''}{2Q}+\frac{\theta\dot{Q}F'}{2Q^2}\right]\mathcal{W}_a-\left[ \frac{\theta F'}{2Q}\mathcal -\theta F''\right] \mathcal{L}_a.\\&&
		\dot{\mathcal W}_a=\mathcal{L}-\frac{w \dot{Q}}{1+w}D^m_a,\\&&	
		\dot{\mathcal L}_a=\frac{\dddot{Q}}{\dot{Q}}\mathcal W-\frac{w \ddot{Q}}{1+w}D^m_a.
	\end{eqnarray}
	By appling the scalar and harmonic decomposition technique as \cite{38, ntahompagaze2018study,sami2021covariant,sahlu2020scalar,ntahompagaze2020multifluid}, the second-order evolution equaion in $f(Q)$ gravity is yield as
	\begin{eqnarray}
		&&\ddot{\Delta}_m^k= -\left[ \frac{2\theta}{3}+\frac{\dot{Q}F'}{2Q}+\dot{Q}F''-w\theta\right]\dot{\Delta}_m^k-\left[-w F-\theta\dot{Q}F''w+\frac{\theta\dot{Q}F'w}{2Q},\nonumber\right.\\&&\left.-\frac{(1+3w)\rho_m}{2}(1-w)-\tilde{\nabla}^2 w\right]\vartriangle_m^k-\left[ \frac{1}{2}F'   -QF''-F'+\theta\dot{Q}F''',\nonumber\right.\\&&\left.-\frac{\theta\dot{Q}F''}{2Q}+\frac{\theta\dot{Q}F'}{2Q^2}\right](1+w)\mathcal{W}^k+\left[ \frac{\theta F'}{2Q} -\theta F''\right](1+w)\dot{\mathcal{W}}^k. \label{125}\\&&
		\ddot{\mathcal W}^k=\frac{\dddot{Q}}{\dot{Q}}\mathcal W^k-\frac{2w \ddot{Q}}{1+w}\Delta_m-\frac{w \dot{Q}}{1+w}\dot{\Delta}_m^k\;,\label{74}
	\end{eqnarray}
	where 
\begin{equation}\label{wavenode}
 k = \frac{2\pi a}{\lambda}\;,
\end{equation}
$k$ being the comoving wave-number and $\lambda$ the wavelength of the perturbations.
  We apply the quasi- static approximation technique which is the fluctuations of nonmetricity density $\mathcal{L}$ and $\mathcal{W}$ are assumed to be constant with time, i.e. $\dot{\mathcal W}^k=\ddot{\mathcal W}^k=\dot{\mathcal L}^k\approx0$, and the second order scalar linear equation reduced to
		\begin{eqnarray}
		&&\ddot{\vartriangle}_m^k=-\left[ \frac{2\theta}{3}+\frac{\dot{Q}F'}{2Q}-\dot{Q}F''-w\theta+\left( \frac{1}{2}F'   -QF''-F'+\theta\dot{Q}F'''-\frac{\theta\dot{Q}F''}{2Q},\nonumber\right.\right.\\&&\left.\left.+\frac{\theta\dot{Q}F'}{2Q^2}\right) \frac{w \dot{Q}^2}{\dddot{Q}}\right]\dot{\Delta}_m^k-\left[-w F-\theta\dot{Q}F''w+\frac{\theta\dot{Q}F'w}{2Q}-\frac{(1+3w)\rho_m}{2}(1-w),\nonumber\right.\\&& \left.+\frac{k^2}{a^2} w+\left(\frac{1}{2}F'   -QF''-F'+\theta\dot{Q}F'''-\frac{\theta\dot{Q}F''}{2Q}+\frac{\theta\dot{Q}F'}{2Q^2} \right)\frac{2w\dot{Q} \ddot{Q}}{\dddot{Q}}\right]\vartriangle_m^k.\label{665}
	\end{eqnarray}
As mentioned earlier, we admit here the power-law $f(Q) = Q^n$ gravity model for illustrative purpose. We also use the the redshift-space transformation technique as presented in \cite{sahlu2020scalar,sami2021covariant,ntahompagaze2020multifluid}, i.e.,for any time derivative functions $f$ transform into a redshift derivative as $\dot{f} = H\frac{df}{dN}$, where $N = \ln{a}$. Thereofore, the above second order evolution equation \eqref{665} for the $f(Q)$ model reduced to
	\begin{eqnarray}
		&&\Delta''_m= \Bigg[ 2-3w-\frac{3}{2}(1+w)+\frac{\mathcal{G}n(1-\Omega_m)}{(1-n)}-\frac{\mathcal{G}n(1-\Omega_m)}{2}-\Big(\frac{(1-\Omega_m)n}{2(1-n)} +(1-\Omega_m)n \nonumber\\&& -\frac{(1-\Omega_m)}{(1-n)}n-\mathcal{G}(1-\Omega_m)n(n-2)+\frac{\mathcal{G}(1-\Omega_m)n(n-2)}{2}+\frac{\mathcal{G}(1-\Omega_m)n}{2(1-n)} \Bigg)\frac{2w }{3(1+w)^2} \Bigg]\frac{1}{(1+z)}{\Delta}'_m \nonumber\\&& -\Bigg[-w \frac{(1-\Omega_m)6}{(1-n)}+\frac{6\mathcal{G}(1-\Omega_m)}{(1-n)} n w +\frac{3\mathcal{G}(1-\Omega_m)}{(1-n)} n -\frac{(1+3w)3\Omega_m}{2}(1-w)+\frac{k^2}{H^2a^2} w \nonumber\\&& +\Bigg(  \frac{(1-\Omega_m)n}{2(1-n)}   +(1-\Omega_m)n-\frac{(1-\Omega_m)}{(1-n)}n-\mathcal{G}(1-\Omega_m)n(n-2)   +\frac{\mathcal{G}(1-\Omega_m)n}{2} w \nonumber\\&&+\frac{\mathcal{G}(1-\Omega_m)n}{2(1-n)}\Bigg)  \frac{4w }{(1+w)^2} \Bigg]\frac{1}{(1+z)^2}\Delta_m\;, \label{dog}  
	\end{eqnarray}
	where 
	$$\mathcal{G}=\frac{-6-6\Omega_m}{6+6\Omega_mn+12n^2-12n^2\Omega_m}.$$ Eq. \eqref{665} is the a closed system equation and we compute the growth of density contrast from this equation in different approches. For the case of $f(Q)= Q$,  equation \eqref{665} reduced to $\Lambda CDM$ model as presented in \cite{sahlu2020scalar,sami2021covariant,ntahompagaze2020multifluid}. 
	We admit the defined normalized energy density contrast for matter fluid  as presented in \cite{sahlu2020scalar} as
	\begin{equation}
		\delta^k(z)=\frac{\Delta_m^k(z)}{\Delta_{in}}.
	\end{equation}
	where the subscript in refers initial value of $\Delta_m (z)$ at redshift $z_{in}$. For GR also we have
	\begin{equation}
		\delta_{GR}=\frac{\Delta_m^k(z)(n=0)}{\Delta_{in}(z_{in})}.
	\end{equation}
	The normalized energy density  can be defined as presented in \cite{sahlu2020scalar}
 \begin{equation}
 \delta_{[m,ch]}(z)=\frac{\Delta _{[m,ch]}(z)}{\Delta (z_{in})} \label{normalizeddensity} \;,                                                                                                                     
 \end{equation}
where $\Delta_{in}$ is the initial value of $ \Delta_{[m,ch]}(z)$ at $z_{in} = 1100$, since the variation of CMB temperature detected observational in the order of $10^{-3}$ \cite{smoot1992structure} at $z\approx 1100$.
		\section{The growth density contrast in $f(Q)$ gravity}
		In this section, we  are going to present the solutions and numerical results of the density contrast ($\delta(z)$) from Eq. \eqref{dog} in dust ($w =0$) and  radiation ($w= \frac{1}{3}$) dominanted univerese.		
  \subsection{Dust Dominated Universe}
\begin{figure}[ht]
  	\centering
  	\includegraphics[width=10 cm, height=8cm]{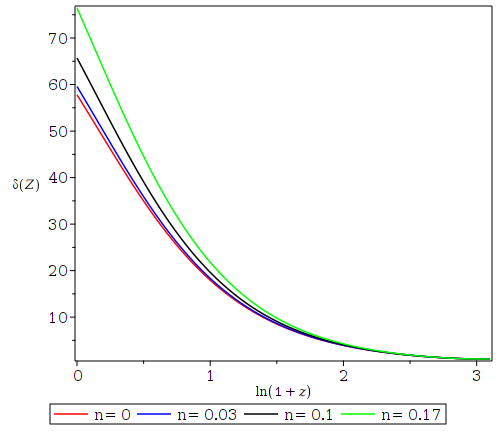}
  	\caption{The growth of matter contrasts in the dust dominated universe in $f(Q)$ gravity.}
  	\label{fig:figuresdust}
  \end{figure}

   The evolution equation in the presence of dust fluid and nonmetricity scalar as
  \begin{eqnarray}
  	&&{\Delta}''_d=
  	\frac{1}{1+z}\left(\frac{1}{2}+\frac{\mathcal{G}n(1-\Omega_m)}{(1-n)}-\frac{\mathcal{G}n(1-\Omega_m)}{2} \right){\Delta'}_d +\frac{3\Omega_m}{2(1+z)^2} \Delta_d. 
  \end{eqnarray}
  The solutions of the above Eq. \eqref{162} are yield as
  \begin{equation}
   \Delta(z) = C_1 \left(1+z \right)^{\frac{\alpha}{2}+\frac{1}{2}+\frac{\sqrt{\alpha^{2}+2 \alpha +4 \beta +1}}{2}}+{C_2} \left(1+z \right)^{\frac{\alpha}{2}+\frac{1}{2}-\frac{\sqrt{\alpha^{2}+2 \alpha +4 \beta +1}}{2}} \label{162}
  \end{equation}
where $$\alpha = \frac{1}{2}+\frac{G n \left(1-\Omega \right)}{1-n}+\frac{G n \left(1-\Omega \right)}{2}\qquad\mbox{and} \qquad \beta = \frac{2}{3}$$ 
  The numerical results of  Eq. \eqref{162}  is presented in  Fig. \ref{fig:figuresdust}. For illustrative purpose, we use the constrained valuse of $n = 1.0^{+07}_{-07}$ from  Fig. \ref{fig:f_Q}. From the plots at $n = 0$, the results exactly reduced to GR limit as presented in \cite{sahlu2020scalar}.
  \subsection{Radiation Dominated Universe}
  The evolution equation in the presence of radiation fluid and nonmetricity scalar as
  \begin{eqnarray}
		&&\Delta''_m= \Bigg[ -\frac{3}{2}+\frac{\mathcal{G}n(1-\Omega_m)}{(1-n)}-\frac{\mathcal{G}n(1-\Omega_m)}{2}-\Bigg(\frac{(1-\Omega_m)n}{2(1-n)} +(1-\Omega_m)n \nonumber\\&& -\frac{(1-\Omega_m)}{(1-n)}n-\mathcal{G}(1-\Omega_m)n(n-2)+\frac{\mathcal{G}(1-\Omega_m)n(n-2)}{2}+\frac{\mathcal{G}(1-\Omega_m)n}{2(1-n)} \Bigg)\frac{1}{8} \Bigg]\frac{1}{(1+z)}{\Delta}'_m \nonumber\\&& -\Bigg[- \frac{(1-\Omega_m)6}{3(1-n)}+\frac{6\mathcal{G}(1-\Omega_m)}{3(1-n)} n  +\frac{3\mathcal{G}(1-\Omega_m)}{(1-n)} n -\frac{4\Omega_m}{3}+\frac{k^2}{3H^2a^2}  \nonumber\\&& +\Bigg(  \frac{(1-\Omega_m)n}{2(1-n)}   +(1-\Omega_m)n-\frac{(1-\Omega_m)}{(1-n)}n-\mathcal{G}(1-\Omega_m)n(n-2)   +\frac{\mathcal{G}(1-\Omega_m)n}{6}  \nonumber\\&&+\frac{\mathcal{G}(1-\Omega_m)n}{2(1-n)}\Bigg)  \frac{3 }{4} \Bigg]\frac{1}{(1+z)^2}\Delta_m\;. \label{dog}  
	\end{eqnarray}
	\subsection{Long Wavelength Mode}
\begin{figure}[ht]
		\centering
	\includegraphics[width=10cm, height=5cm]{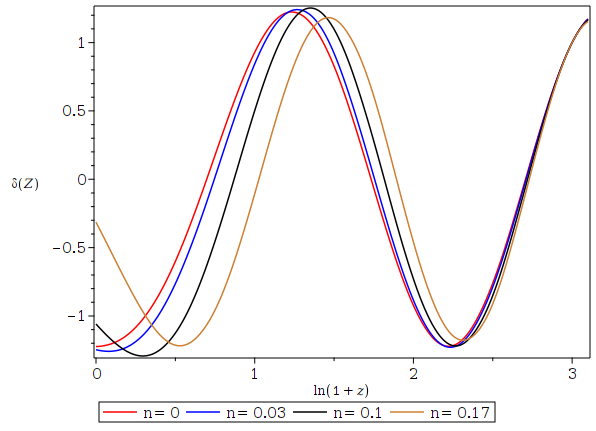}
	\caption{The growth of energy contrasts in the radiation dominated universe in longwave length mode for different values of $n$.}
	\label{fig:radiationlongwavelength}
	\end{figure}

For the case of longwave length, $\frac{k^2}{3a^2H^2} << 1$, all cosmological fluctuations begin and remain inside the Hubble horizon. The solutions of Eq. \eqref{dog} in longwave mode yield as
\begin{eqnarray}
 \Delta_{r} = C_1 \left(1+z \right)^{-\frac{l}{16}+\frac{G n \left(1-\Omega \right)}{2-2 n}-\frac{G n \left(1-\Omega \right)}{4}+\Gamma}+C_2 \left(1+z \right)^{-\frac{l}{16}+\frac{G n \left(1-\Omega \right)}{2-2 n}-\frac{G n \left(1-\Omega \right)}{4}-\Gamma} \label{YT}
\end{eqnarray}
where 
\begin{eqnarray*}
&& \Gamma = \frac{\sqrt{\Psi}}{16}\;,\\&&
\Psi = l^{2}-16 l \left(-1+\frac{G n \left(1-\Omega \right)}{1-n}-\frac{G n \left(1-\Omega \right)}{2}\right)+64 \left(-1+\frac{G n \left(1-\Omega \right)}{1-n}-\frac{G n \left(1-\Omega \right)}{2}\right)^{2}\\&& -256 \eta -16 l -64+\frac{128 G n \left(1-\Omega \right)}{1-n}-64 G n \left(1-\Omega \right)-192 \zeta\\&& \eta = -2 \Omega -\frac{2 \left(1-\Omega \right)}{1-n}+\frac{5 G n \left(1-\Omega \right)}{1-n}\;,\\&&l = \frac{n \left(1-\Omega \right)}{2-2 n}+n \left(1-\Omega \right)-\frac{\left(1-\Omega \right) n}{1-n}-\frac{G \left(1-\Omega \right) n \left(n -2\right)}{2}+\frac{G n \left(1-\Omega \right)}{2-2 n}\;,\\&& \zeta = \frac{n \left(1-\Omega \right)}{2-2 n}+n \left(1-\Omega \right)-\frac{\left(1-\Omega \right) n}{1-n}-G \! \left(1-\Omega \right) n \left(n -2\right)+\frac{G n \left(1-\Omega \right)}{2}+\frac{G n \left(1-\Omega \right)}{2-2 n}\;. 
\end{eqnarray*}
Then the numerical results os density contrast for longwave mode is presented in Fig Eq. \ref{fig:radiationlongwavelength} from Eq. \eqref{YT}. From the plots we observe the amplitude of density contrast is decaying through redshift. 
	\subsection{Short Wavelength Mode}
\begin{figure}[ht]
		\begin{minipage}{0.49\textwidth}
		\includegraphics[width=8cm, height=5cm]{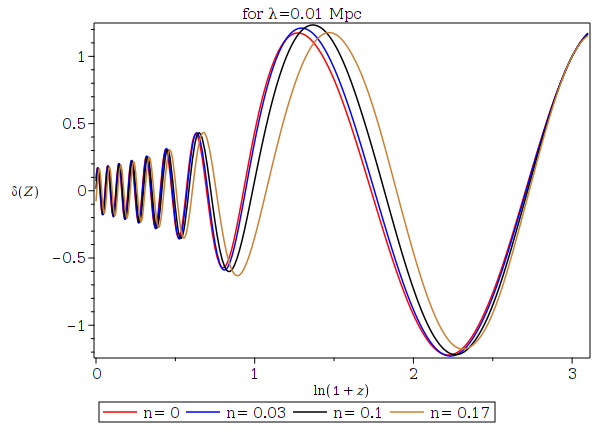}
		\caption{The growth of matter contrasts in the radiation dominated universe in short wave length.}
		\label{fig:radiationshortwavelength}
		 \end{minipage}
	\qquad
\begin{minipage}{0.49\textwidth}
		\includegraphics[width=8cm, height=5cm]{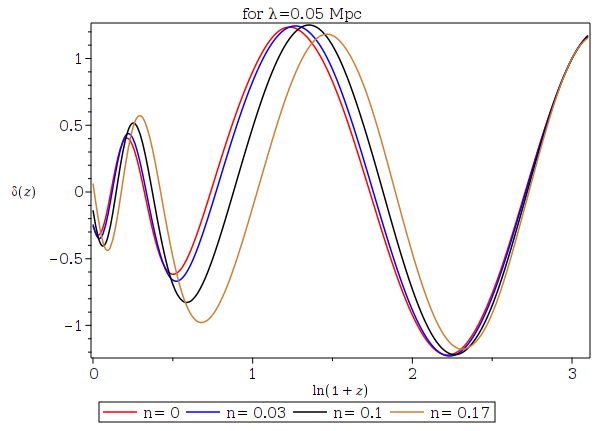}
		\caption{The growth of matter contrasts in the radiation dominated universe in short wave length.}
		\label{fig:radiationshortwavelength1}
	\end{minipage}
	\end{figure}

	For the case of short wavelength mode $\frac{k^2}{3a^2H^2} >> 1$, the solutions of Eq. \eqref{dog} yield as
	\begin{eqnarray}
&& \Delta_r(z) = C_3\left(1+z \right)^{\frac{1}{2}+\frac{\alpha}{2}} BesselJ\Bigg({\frac{\sqrt{\alpha^{2}+2 \alpha +4 \beta +1}}{4}}, \frac{\sqrt{3}\, \pi}{3 \lambda  \left(1+z \right)^{2}}\Bigg)\\&&+C_4 \left(1+z \right)^{\frac{1}{2}+\frac{\alpha}{2}} BesselY \left({\frac{\sqrt{\alpha^{2}+2 \alpha +4 \beta +1}}{4}}, \frac{\sqrt{3}\, \pi}{3 \lambda  \left(1+z \right)^{2}}\right)\nonumber
	\end{eqnarray}
From Eq.\eqref{wavenode}, we rewrite as $\frac{k^2}{3a^2H^2}=\frac{16\pi^2}{3\lambda^2(1+z)^4}$ as presented in \cite{sahlu2020scalar}, and the numerical results of the above solutions are presented in Figs. \ref{fig:radiationshortwavelength} and \ref{fig:radiationshortwavelength1} for different values of wavelength $\lambda$. From the plot. we can see the amplitude of density contrast is  oscillates in the short wavelength. for different values of $n$. 
	\newpage
	\section{Summary}
	In this work, we considered the symmetric teleparallelism, it is curvatureless and torsion-free gravity and fully described by nonmetricity scalar $Q$ and  studied the modified gravity  in the regime of most general class of symmetric teleparallel gravity theories whose action is quadratic in the nonmetricity tensor. The analysis is done by using linear cosmological perturbations using 1+3 covariant gauge-invariant formalism. We have also considered the power law $f(Q)$ gravity model  $f(Q)=Q+\alpha  Q^n$.  In which the presented $f(Q)$ gravity model approach provides gravitational alternatives to the dark energy to explain the late and early phase of universe. For both regimes, the symmetric teleparallel gravity is reduced to GR at $f(Q) = Q$ gravity model. 
	\section*{References}
	\bibliographystyle{ieeetr}

\end{document}